# Large-angle quasi-self-collimation effect in a rod-type photonic crystal

**Ming Li,**[1,2] **Wei Li,**[1*] **Haiyang Huang,**[1,2] **Jing Wang,**[1,2] **You Li,**[1,2] **Aimin Wu,**[1] **Zhen Sheng,**[1] **Xi Wang,**[1] **Shichang Zou,**[1] **Fuwan Gan**[1*]

[1]State Key Laboratory of Functional Materials for Informatics, Shanghai Institute of Microsystem and Information Technology, Shanghai, China, 200050
[2]University of Chinese Academy of Science, Beijing, China, 100049
*Corresponding authors: waylee@mail.sim.ac.cn (Wei Li), and fuwan@mail.sim.ac.cn (Fuwan Gan)

**Abstract:** A rod-type photonic crystal (PC) with a rectangular lattice shows a large-angle quasi-self-collimation (quasi-SC) effect by changing the symmetry of its rectangular lattice to straighten one of the isofrequency contours. To investigate the straightness of the isofrequency contour as well as the quasi-SC effect, we propose a straightness factor L based on the method of least squares. With $L \leq L_0$ ($L_0 = 0.01$ is the critical value), the isofrequency contour is sufficiently straight to induce quasi-SC effect with the beam quasi-collimating in the structure. Furthermore, the efficiency of light coupling to the quasi-SC PC is studied, and can be greatly improved by applying a carefully designed antireflection layer. This quasi-SC effect of the PC as well as the coupling structure may see application in novel optical devices and photonic circuits.

## 1. Introduction

During the past two decades, photonic crystals (PCs) have attracted much attention due to their potential application in the miniaturization and integration of optical devices[1]. They exhibit a variety of new physical phenomena, including a photonic bandgap (PBG)[2], the ability to act as a superprism[3], negative refraction[4] and self-collimation (SC)[5]. In an SC PC, the light beams propagate without diffraction, since their propagation directions are forced to be parallel to the group velocity, i.e., $V_g=\nabla_k\omega(\boldsymbol{k})$ where $\omega$ is the optical frequency for a given wave vector $\boldsymbol{k}$. Therefore, the SC effect can be attributed to the flat part of the equi-frequency contours (EFCs)[6, 7]. The SC effect can be used to design a variety of novel SC-based photonic devices, such as channel-less waveguides[8], or devices for diffraction inhibition[9] and subwavelength focusing or imaging[10].

However, there are still a few issues that need to be addressed. For instance, the SC effect might be limited by the light incident angle, making it difficult for super-integrated devices based on the SC phenomenon. In this paper, we will present a model for a large-angle quasi-SC PC, and lower the reflections for our quasi-SC PC by utilizing the destructive interference-based method.

## 2. Model

For this study, we consider a two-dimensional (2D) rod-type silicon PC with a rectangular lattice in the air, as shown on the inset image in Fig. 1(a). The length and breadth of the rectangular lattice are denoted by *b* and *a*, respectively, and the radius of the rod is *r=0.3a*. The silicon rods are assumed to be lossless and nondispersive near the telecom frequency and exhibit a refractive index of 3.5. Utilizing the plane-wave expansion method, the EFCs are then calculated. For simplicity, throughout this paper, only one single frequency *f=0.2 c/a* is considered for the transverse electric (TE) modes. This frequency is located in the first photonic band, and the corresponding isofrequency contour plot is shown in Fig. 1.

# 3. Results and discussion
## 3.1. Straightness of the EFCs and collimation ability of the beams

As shown in Fig. 1(a) to (d), by changing the symmetry of the PC, i.e., by increasing the length–breadth ratio $\beta=b/a$ of the rectangular lattice, a large-angle self-collimation effect can be eventually obtained for the PC. The large-angle effect is indicated by the straightness of the EFC which is plotted over the whole Brillouin zone in Fig. 1(d). As $\beta$ increases, the light beams can reduce the beam divergence, resulting in a self-collimating propagation along the $\Gamma X$ direction. Therefore, in order to improve the collimation ability, a higher straightness of the EFCs must be obtained. As illustrated by Fig. 1, the straightness quality of the EFCs depends on $\beta$.

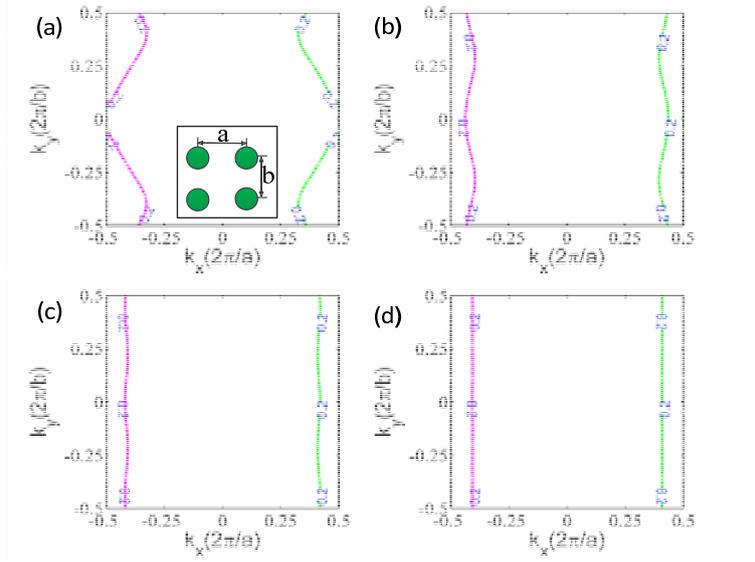

Fig. 1. EFCs for the normalized frequency $f=0.2c/a$, fixed $a$, $r=0.3a$ and different lattice length–breadth ratios $\beta$: (a) $\beta=1.25$, (b) $\beta=1.75$, (c) $\beta=2.3$, (d) $\beta=3.5$. Inset image in (a): schematic illustration of the 2D PC.

To investigate the correlation between length-breadth ratio $\beta$ and EFC straightness, we use the method of least squares[11] to quantify the EFC straightness, as shown in Fig. 1. According to the method of least squares, considering a quasi-straight EFC whose curve can be represented by a function $Y=F(X)$ ($Y$ and $X$ representing $k_y$ and $k_x$ respectively), we assume that a straight line with the equation

$$\overline{Y} = AX + B \tag{1}$$

can be used to fit the quasi-straight EFC, where A and B are undetermined coefficients. In order to obtain A and B, we define $\varepsilon = \sum_{i=1}^{n}\left[Y_i - (AX_i + B)\right]^2$, where $i$ is the number of sampling points, then, by utilizing the minimum condition, i.e., $\partial\varepsilon/\partial A$, $\partial\varepsilon/\partial B$, we obtain

$$A\sum X_i^2 + B\sum X_i = \sum X_i Y_i \tag{2}$$

and

$$A\sum X_i + nB = \sum Y_i \tag{3}$$

From Eq. (2) and Eq. (3), $A$ and $B$ can be calculated as follows:

$$A = \frac{n\sum X_i Y_i - \sum X_i \sum Y_i}{n\sum X_i^2 - (\sum X_i)^2} \quad B = \frac{\sum Y_i \sum X_i^2 - \sum X_i \sum X_i Y_i}{n\sum X_i^2 - (\sum X_i)^2} \quad (4)$$

Finally, we can quantify the straightness quality of the EFCs via the straightness factor

$$L = \Delta L_{max} - \Delta L_{min} \quad (5)$$

Where $\Delta L_{max} = [Y - \overline{Y}]_{max} = [Y - AX - B]_{max}$ and $\Delta L_{min} = [Y - \overline{Y}]_{min} = [Y - AX - B]_{min}$ are the maximum and the minimum convexities, respectively.

Physically, the straightness factor L as defined by Eq. (5) can be used to describe the averaged deviation angle of the propagation direction of a quasi-collimated light beam. For a quasi-collimated light beam, the beam width will become broader when the beam is propagating through the quasi-SC PC, with the waist of beam given by $W(D) \approx W_0 + L\frac{\lambda D}{\pi W_0}$ where $\lambda$ is the wavelength in the air, $W_0$ and $D$ (we assume $D \gg \lambda$, $D \gg W_0$) represent the initial waist and propagation distance of the beam, respectively. A smaller $L$ corresponds to a straighter EFC and better self-collimation behavior. Only for $L=0$ a perfectly straight EFC is obtained that corresponds to the strict self-collimation phenomenon without beam divergence, i.e., $W(D)=W_0$ for any $D$.

For most practical applications, the condition $L=0$ is too strict. A sufficiently small L is usually acceptable. We suggest $L_0=0.01$ as the critical value for the straightness factor (corresponding to the pink dash line in Fig. 2). In this case, the quasi-collimated beam does hardly show any diffraction if $L \leq L_0$, e.g., with respect to a typical Gaussian beam with an initial waist $W_0=10\lambda$, the beam broadening is smaller than 1% as the beam propagates over a distance of 100λ. We suggest that such a quasi-collimated beam is sufficient for most practical applications with the self-collimation phenomenon.

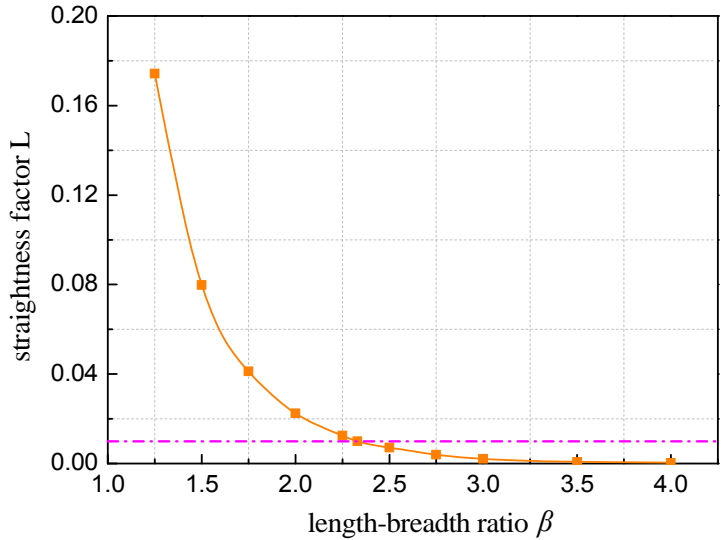

Fig. 2. The straightness factor L plotted as a function of the length–breadth ratio $\beta$ of the rectangular lattice

Using Eq. (5), the straightness factor L of the EFCs can be calculated. The relation between the EFCs' straightness factor and the rectangular lattice structure is shown in Fig. 2. Apparently, the straightness factor L decreases as $\beta$ increases. This result is in good agreement with the results presented in Fig. 1. In Fig. 1(a) to (d), the corresponding straightness factors are *L=0.1743, L=0.041, L=0.01* and *L=8.09×10$^{-4}$*, respectively. In Fig. 2, the condition *L=L$_0$* corresponds to *β=2.3*, with the corresponding EFC shown in Fig. 1(c).

The quasi-collimation phenomenon of the PC at the frequency *f=0.2c/a*, with *β=2.3, a=0.4μm* and *r=0.3a*, is demonstrated in Fig. 3 for different incident angles. The electromagnetic field is calculated by applying the finite differential time domain method (FDTD)[12, 13]. As shown in Fig. 3, the Gaussian beam travels without beam divergence in the quasi-SC PC, and thus the structure exhibits a quasi-SC effect even at a large incident angle.

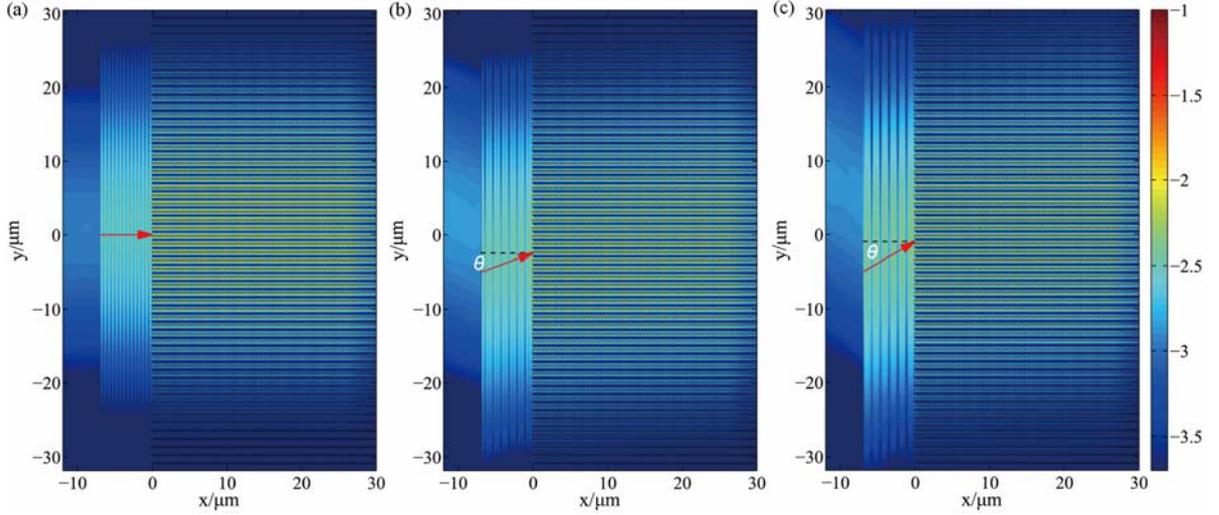

Fig. 3. Field distribution obtained for the illumination of a Gaussian beam into a quasi-SC PC with different incident angles *θ*: (a) *θ* = 0°, (b) *θ* = 20°, (c) *θ* = 30°. The field distribution is displayed using a logarithmic color map.

### 3.2 Coupling analysis

In this section, the efficiency of light coupling to the quasi-SC PC is studied. As shown in Fig. 3, large coupling losses (about 50%) occur due to the strong reflection. To improve the coupling efficiency, a destructive interference-based method is presented. As illustrated in Fig. 4, an antireflection layer (ARL) is applied in front of quasi-SC PC. The ARL also consists of an array of silicon rods with the lattice constant *b=2.3a*, which is equal to the length of the quasi-SC PC's rectangular lattice. The radius of the silicon rods in the ARL is $r_{arl}$, and the distance between the ARL and the quasi-SC PC is $d_{arl}$. The parameters $d_{arl}$ and $r_{arl}$ should be well-designed to improve the coupling efficiency.

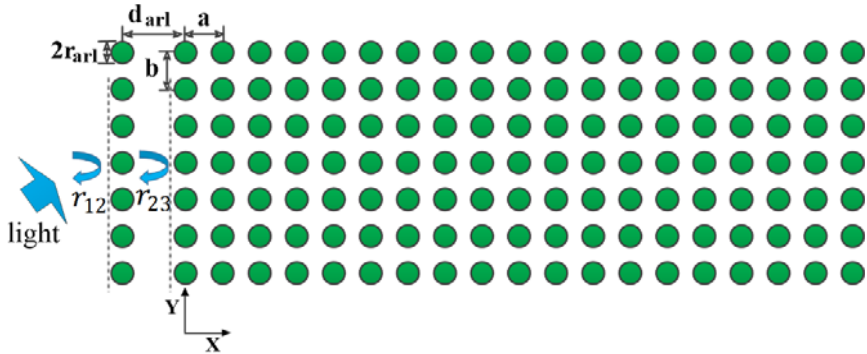

Fig. 4. Schematic illustration of the ARL placed in front of the quasi-SC PC. The variable ARL parameters are the distance $d_{arl}$ and the radius $r_{arl}$.

The coupling efficiency $\kappa=1-|r_{tot}|^2$ is directly determined by the total reflection coefficient $r_{tot}$, which can be written as:

$$r_{tot} = \frac{r_{12} + r_{23}e^{i2\beta}}{1 + r_{12}r_{23}e^{i2\beta}} \qquad (6)$$

where $\beta = k_0 \cdot (d_{arl} - r - r_{arl})$ is the phase shift of the light beam as it crosses the ARL ($k_0$ denotes the wave vector in the air), $r_{12}$ is the reflection coefficient of the ARL in the air and $r_{23}$ is the reflection coefficient of the semi-infinite quasi-SC PC when the light beam is hitting the PC surface from air. Both $r_{12}$ and $r_{23}$ can be calculated by applying the multiple scattering method[14].

Now, we try to find suitable $r_{arl}$ and $d_{arl}$ values in order to reduce the total reflection $|r_{tot}|^2$. First, the coupling efficiency for an initial value of $r_{arl}$ with $r_{arl}=r$ is calculated, and the resulting color map is shown in Fig. 5(a). In this figure, the red region corresponds to a high coupling efficiency. As shown in Fig. 5(a), there are two regions that allow for a high coupling efficiency for a relatively wide range of incident angles. One can choose suitable $d_{arl}$ according to Fig. 5(a), e.g., for a $d_{arl}$ of approx. *3.28a* for incident angle 0° to 20°, or *3.44a* for incident angle 20° to 30°. Then, for the optimized $d_{arl}$, one can further improve the coupling efficiency by optimizing $r_{arl}$, as shown in Fig. 5(b) and Fig. 5(c). In Fig. 5(b) and Fig. 5(c), the values of $d_{arl}$ are fixed to be *3.28a* and *3.44a*, respectively. The enhanced coupling efficiency versus incident angle is shown in Fig. 5(d). From Fig. 5(d), we can see by using the well-designed ARL, the coupling efficiency is greatly improved for a wide incident angle range.

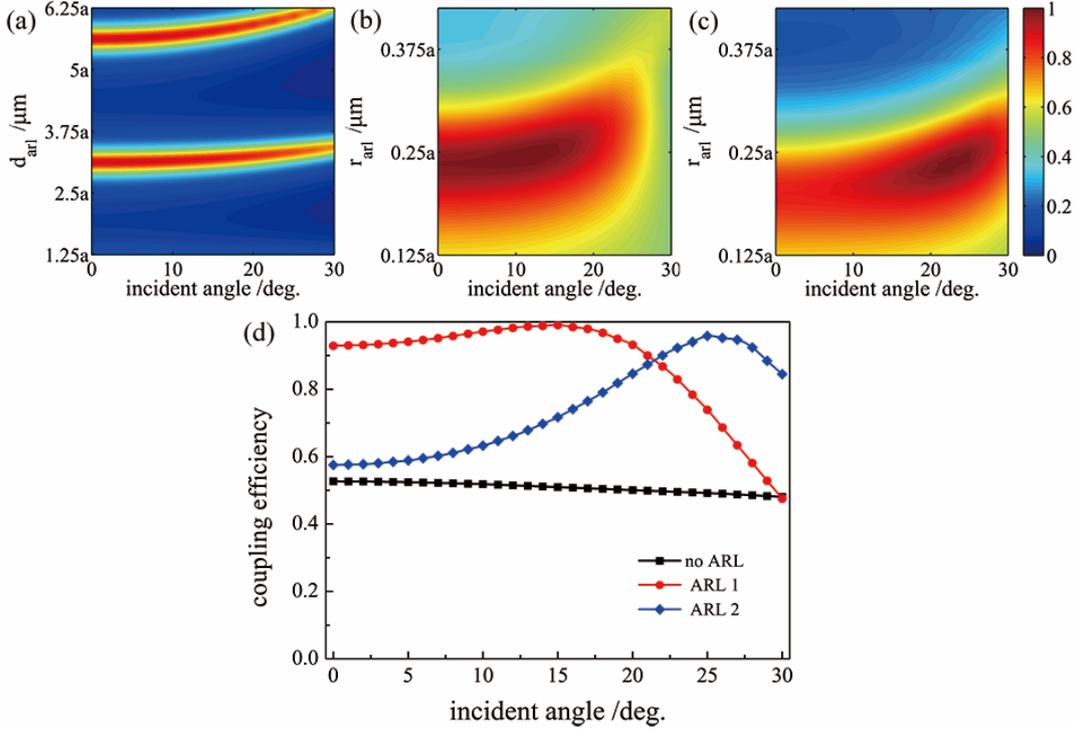

Fig. 5. (a) Coupling efficiency as a function of $d_{arl}$ and the incident angle for $r_{arl}$ fixed at *0.3a.* (b) Coupling efficiency as a function of $r_{arl}$ and an incident angle θ for an optimized $d_{arl}$ =3.28a. (c) Coupling efficiency as a function of $r_{arl}$ and an incident angle θ for an optimized $d_{arl}$ =3.44a. (d) Coupling efficiency as a function of the incident angle without and with ARL. The structure parameters of ARL 1 and ARL 2 are $d_{arl}$ = 3.28a, $r_{arl}$ = 0.26a, and $d_{arl}$ = 3.44a, $r_{arl}$ = 0.26a, respectively.

To validate the improvement of the coupling efficiency, we calculated the field distribution for two Gaussian beams launching on the quasi-SC PC with an ARL, and the resulting distributions are shown in Fig. 6. For the optimized ARL structure, the Gaussian beams almost perfectly couple into the quasi-SC PC with almost no reflection. Comparing Fig. 6 with Fig. 3, the reflection is suppressed by the optimized ARL.

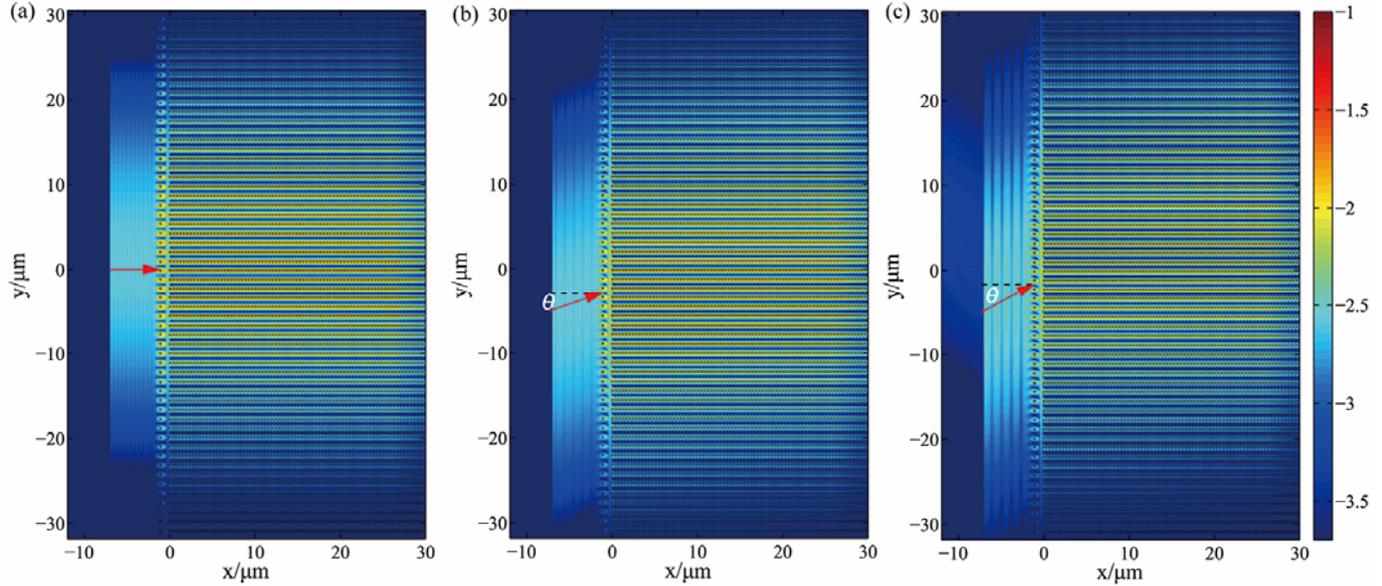

Fig. 6. Field distribution with $r_{arl}$ and $d_{arl}$ optimized for different beam incident angles $\theta$: (a) $\theta = 0°$, with ARL 1. (b) $\theta = 20°$, with ARL 1. (c) $\theta = 30°$, with ARL 2. The parameters of ARL1 and ARL 2 are identical to Fig.5, and the field distribution is displayed using a logarithmic color map.

## 4. Conclusion

In this paper, a quasi-self-collimation effect is obtained by changing the symmetry of a rectangular-lattice photonic crystal. The quasi-self-collimation effect is quantified by a straightness factor L that is based on the method of least squares. When the straightness factor L decreases, the photonic crystal possesses a more powerful self-collimation effect. Besides, the efficiency of light coupling to the quasi-SC PC is investigated, and is greatly improved by applying a carefully designed antireflection layer. This quasi-SC effect of the PC as well as the coupling structure may see applications in novel optical devices and photonic circuits.

## Acknowledgements

This research was sponsored by the National Natural Science Foundation of China (Grant Nos. 61475180, 61107031, 11204340, and 61275112), the National High Technology Research and Development Program of China (Grant No.2012AA012202), and the Science and Technology Commission of Shanghai Municipality (Grant Nos. 14JC1407601, and 14JC1407602).